\documentclass[reprint]{revtex4-1}
\pdfoutput=1

\usepackage{graphicx}
\usepackage{amsmath}
\usepackage{amsfonts}
\usepackage{amssymb}
\usepackage{amsthm}
\usepackage{mathrsfs}
\usepackage[retainorgcmds]{IEEEtrantools}
\usepackage{tikz}
\usepackage{setspace}
\usepackage{subfig}
\usepackage{hyperref}
\input{Qcircuit}

\begin{document}

\title{Scale invariance and efficient classical simulation of the quantum Fourier transform}
\author{Kieran J. Woolfe}
\author{Charles D. Hill}
\author{Lloyd C. L. Hollenberg}
\affiliation{Center for Quantum Computation and Communication Technology, School of Physics, University of Melbourne, Victoria, 3010, Australia}

\begin{abstract}
We provide numerical evidence that the quantum Fourier transform can be efficiently represented in a matrix product operator with a size growing relatively slowly with the number of qubits.
Additionally, we numerically show that the tensors in the operator converge to a common tensor as the number of qubits in the transform increases.
Together these results imply that the application of the quantum Fourier transform to a matrix product state with $n$ qubits of maximum Schmidt rank $\chi$ can be simulated in $O(n \left(\log (n)\right)^2 \chi^2)$ time.
We perform such simulations and quantify the error involved in representing the transform as a matrix product operator and simulating the quantum Fourier transform of periodic states.
\end{abstract}

\maketitle

\section{Introduction}
\label{sec:int}
A central problem of quantum computing is determining the origin and nature of the speedup provided over classical computing.
One approach to this problem is to study which classes of quantum computations can be simulated efficiently by classical means.
Such computations must be missing a central feature of quantum computing, separating it from the classical counterpart.
There have been many results in this area.
These include the Gottesman-Knill theorem \cite{Gottesman1998, Nest2009}, which states that circuits composed only of Clifford group gates can be efficiently simulated classically.
Other results include the efficient classical simulation of  match-gate circuits \cite{Valiant2001, Jozsa2008}, circuits which generate limited entanglement \cite{Vidal2004, Jozsa2003}, circuits whose graph representation has restricted topological properties \cite{Markov2008, Yoran2006, Jozsa2006} and circuits with sparse output distributions \cite{Schwarz}.

The quantum Fourier transform (QFT) is an important part of several quantum algorithms, including quantum simulation \cite{Lloyd1996} and Shor's algorithm \cite{shor95pf}.
Each of these provides an exponential speedup over the fastest known classical alternatives.
In our discussion of the QFT we will focus on its role in Shor's algorithm.
The QFT is the most intuitively quantum mechanical part of Shor's algorithm.
That is, it contains Hadamard and controlled phase-rotation gates, neither of which have a classical analogue.
The full QFT does not display any of the features found in previous studies to allow efficient classical simulation.
Despite this, the approximate QFT (AQFT) is efficiently classically simulatable for input states with limited entanglement.
This was first shown in \cite{Aharonov2006, Yoran2007a} using a tensor contraction simulation method. 
Together with results showing that the AQFT is sufficient for many computational tasks including Shor's algorithm \cite{Coppersmith2002, Barenco1996, Fowler2004a, Nam2013}, this result is sufficient to show that the QFT is efficiently classically simulatable to high fidelity for a limited class of input states.
It was shown in \cite{Browne2007} using matrix product states that a terminating QFT can be efficiently simulated for any input state with limited entanglement.
Additionally, in \cite{Abbott2012} a classical algorithm to obtain the results of the QFT on separable states is derived.
That this algorithm is simpler than the QFT suggests that the quantum speedup of the QFT lies in the quantum parallelism of its input state rather than its innate complexity.

In this paper, we use matrix product operators to simulate the quantum Fourier transform.
Our numerical results imply that for weakly entangled input states over $n$ qubits, the resources required for this simulation scale as $O(n \left(\log(n)\right)^2)$, a significant improvement on earlier results.
The simulation tools we use are also more straightforward than those found in \cite{Aharonov2006, Yoran2007a} and have a wider applicability than other results concerning the classical simulatability of the QFT.
In section \ref{sec:mps} we will review matrix product states and matrix product operators.
Section \ref{sec:qft} discusses their use to simulate the QFT.
In section \ref{sec:trunerror} we will present numerical results showing the errors associated with such simulation to be minimal.
Finally we will discuss the implications of this work and how it is related to the computational speedup of Shor's algorithm in section \ref{sec:disc}.

\section{Matrix product states and operators}
\label{sec:mps}
A matrix product states (MPS) of $n$ qubits is a quantum states with the form \cite{Fannes1992, Vidal2003}:
\begin{align}
  \ket{\psi} = & \sum_{i_1,i_2,\ldots,i_n} \sum_{\alpha_1, \alpha_2, \ldots, \alpha_n} \Gamma^{[1]\alpha_1}_{i_1} \lambda^{[1]}_{\alpha_1} \Gamma^{[2]\alpha_1 \alpha_2}_{i_2} \ldots \nonumber \\ & \quad \Gamma^{[n-1]\alpha_{n-2} \alpha_{n-1}}_{i_{n-1}} \lambda^{[n-1]}_{\alpha_{n-1}} \Gamma^{[n]\alpha_{n-1}}_{i_n} \ket{i_1} \ldots \ket{i_n},
\end{align}
where $\ket{i_j}$ is the state of the $j$th qubit in the system.
The determination of a coefficient of this state requires the contraction of a series of two-dimensional tensors $\{\Gamma^{[j]\alpha_{j-1} \alpha_j}_{i_j} \}$ and one dimensional vectors $\lambda^{[j]}_{\alpha_j}$ where the $i_j$ are set to specify the coefficient required in the computational basis.
The $\alpha_i$ are ancillary indices and will henceforth be referred to as bonds between different parts of the system.
The connectivity of the tensors reflects a one-dimensional ordering of the qubits in a state.

In the canonical form of a MPS, the decomposition from a state vector to matrix product form is accomplished by a series of singular value decompositions, in which the $\Gamma$ tensors are unitary and the $\lambda$ vectors contain the singular values.
Each bond joins two tensors and corresponds to a bipartition of the state.
The rank of a bond is the Schmidt rank of this bipartition.
A state with little entanglement will have low Schmidt ranks and so the tensors used to encode the state in a MPS will be small.
As such, it is possible to efficiently simulate quantum states with low entanglement with a MPS \cite{Vidal2003, Vidal2004}.
The MPS form is also useful because it allows the truncation of the number of singular values in each bipartition.
As the singular values are ordered, it is straightforward to remove the smallest ones and then to renormalise the state.

It is possible to generalise the structure of matrix product states in several ways.
A simple generalisation is to keep the linear connectivity of the tensor network but to encode an operator rather than a state.
\begin{align}
  \mathbf{O} = & \sum_{i_1,\ldots,i_n} \sum_{j_1,\ldots,j_n} \sum_{\alpha_1, \ldots, \alpha_{n-1}} O^{[1]\alpha_1}_{i_1 j_1} \gamma^{[1]}_{\alpha_1} O^{[2]\alpha_1 \alpha_2}_{i_2 j_2} \ldots \nonumber\\
  & O^{[n-1]\alpha_{n-2} \alpha_{n-1}}_{i_{n-1} j_{n-1}} \gamma^{[n-1]}_{\alpha_{n-1}} O^{[n]\alpha_{n-1}}_{i_n j_n} \ket{i_1} \ldots \ket{i_n}\bra{j_1} \ldots \bra{j_n}. \nonumber
\end{align}
This is called a Matrix Product Operator (MPO) and was introduced in \cite{Verstraete2004a, Zwolak2004}.
Similarly to MPSs, the canonical form of a MPO is created with a series of singular value decompositions.
The bond ranks of the MPO are then the Schmidt numbers of the operator.
The maximal bond dimension of a MPO thus gives the Hartley strength of the operator \cite{Nielsen2003}.

The Schmidt rank of a state is a measure of the amount of entanglement in the state.
However, the Schmidt number of an operator does not have as straight-forward an interpretation.
An operator with a high Schmidt number may have a high amount of classical correlating power but little entangling ability.
This is illustrated in the case of two-qubit unitary operations by the SWAP gate, which has the Schmidt-operator decomposition $1/2 (I \otimes I + X \otimes X + Y \otimes Y + Z \otimes Z)$. 
The Schmidt number of four is the maximum possible for a two-qubit operator, but the gate has no entangling power.

Applying a MPO to a MPS produces a new MPS in which each tensor is the product of a tensor in the original MPS and a tensor in the MPO, each representing the same qubit:
\begin{align}
  \label{eqn:mpo}
  \ket{\psi'} = &\sum_{i_1,i_2,\ldots,i_n} \sum_{\beta_1, \beta_2, \ldots, \beta_n} \Gamma^{'[1]\beta_1}_{i_1} \lambda^{'[1]}_{\beta_1} \Gamma^{'[2]\beta_1 \beta_2}_{i_2} \ldots \nonumber\\ &\Gamma^{'[n-1]\beta_{n-2} \beta_{n-1}}_{i_{n-1}} \lambda^{'[n]}_{\beta_{n-1}} \Gamma^{'[n]\beta_{n-1}}_{i_n} \ket{i_1} \ldots \ket{i_n},
\end{align}
where we have relabelled the new ancillary indices to include those from both the state and the operator:
\begin{IEEEeqnarray*}{c}
  \Gamma^{'[l] \beta_{l-1} \beta_l}_{i_l} = \Gamma^{'[l] \alpha_{l-1} \alpha_l \mu_{l-1} \mu_l}_{i_l} 
  = \sum_{j_l} \Gamma^{[l] \alpha_{l-1} \alpha_l}_{j_l} O^{[l] \mu_{l-1} \mu_l}_{j_l i_l},\nonumber\\
 \lambda^{'[l]}_{\beta_l} = \lambda^{'[l]}_{\alpha_l \mu_l} = \lambda^{[l]}_{\alpha_l} \gamma^{[l]}_{\mu_l}.
\end{IEEEeqnarray*}
This multiplication can be followed by a new singular value decomposition at each bipartition to return the state to canonical form.
Before the singular value decomposition, the new state will have a bond rank of $c_j d_j$ at site $j$ where $c_j$ is the bond rank of the MPS and $d_j$ is the bond rank of the MPO.
This rank invites an interpretation of the Schmidt number of an operator as an upper bound for the amount by which the Schmidt rank of a state can increase upon application of the operator.
If the MPO has bond rank $d$ for that partition, the Schmidt rank will be at most multiplied by $d$.
However, in many cases the Schmidt rank after application will be much lower than this.



\section{Simulation of the QFT with a MPO}
\label{sec:qft}

The quantum Fourier transform can be written in operator form:
\begin{align}
  \label{eqn:qft}
  \frac{1}{\sqrt{N}} \sum_{j,k=0}^{N-1} e^{2 \pi i j k / N} \ket{j}\bra{k}.
\end{align}

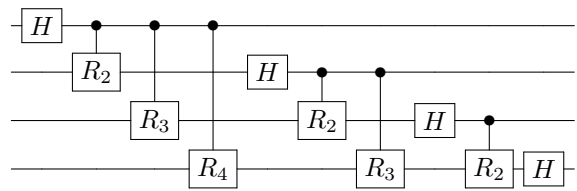
\begin{figure}[hbt]
  \mbox{
  \Qcircuit @C=0.4em @R=0.4em {
    & \gate{H} & \ctrl{1}   & \ctrl{2}   & \ctrl{3}   & \qw      & \qw        & \qw        & \qw      & \qw        & \qw      & \qw\\
    & \qw      & \gate{R_2} & \qw        & \qw        & \gate{H} & \ctrl{1}   & \ctrl{2}   & \qw      & \qw        & \qw      & \qw\\
    & \qw      & \qw        & \gate{R_3} & \qw        & \qw      & \gate{R_2} & \qw        & \gate{H} & \ctrl{1}   & \qw      & \qw\\
    & \qw      & \qw        & \qw        & \gate{R_4} & \qw      & \qw        & \gate{R_3} & \qw      & \gate{R_2} & \gate{H} & \qw
  }
}
  \caption{The canonical decomposition of the quantum Fourier transform with four qubits.}
  \label{fig:full}
\end{figure}

This equation can be expanded to give output values at individual qubits:
\begin{align}
  \ket{j_1,\ldots,j_n} \rightarrow \frac{1}{2^{n/2}} & \left( \ket{0} + e^{2\pi i 0.j_n} \ket{1} \right) \nonumber \\ & \ldots \left( \ket{0} + e^{2\pi i 0.j_1 \ldots j_{n}} \ket{1} \right),
  \label{eqn:qftl}
\end{align}
where $j = j_1 2^{n-1} + j_2 2^{n-2} + \ldots + j_n 2^0$ and $0.j_l\ldots j_m = j_l/2 + \ldots + j_m/2^{m-l+1}$.
This form of the equation motivates the canonical decomposition of the QFT into quantum gates, which is shown in figure \ref{fig:full}.

The operator-Schmidt decomposition of the QFT has been calculated exactly \cite{Tyson2003} and the maximal Schmidt number of a $n$ qubit transform is $2^n$.
Additionally, all of the singular values are equal.
Simulating the QFT using a MPO representation of (\ref{eqn:qftl}) would thus entail an exponential scaling in terms of execution resources as the number of qubits is increased.

From (\ref{eqn:qftl}) we note that the output at the first qubit depends only upon the input value at the last qubit, the input at the second depends upon the output at the last two qubits and so on.
As such, the operator displays similar classical correlations between the input and output values to those in our earlier swap gate example.
These correlations are expensive to encode in a MPO.
A simple re-ordering of the input or output qubit values (but not both) from equation (\ref{eqn:qftl}) produces a more easily encoded operator:
\begin{align}
  \ket{j_1,\ldots,j_n} \rightarrow \frac{1}{2^{n/2}} & \left( \ket{0} + e^{2\pi i 0.j_1} \ket{1} \right) \nonumber \\ & \ldots \left( \ket{0} + e^{2\pi i 0.j_n \ldots j_1} \ket{1} \right).   \label{eqn:qftlr}
\end{align}
In (\ref{eqn:qftlr}) the output at the first qubit depends only upon the input at the first qubit, the output at the second qubit upon the input at the first two qubits and so on.
This ordering thus requires less information to be communicated across bonds and can be encoded in a smaller MPO.

%
To construct a MPO encoding (\ref{eqn:qftlr}) one can construct a MPO of (\ref{eqn:qftl}) and then apply the SWAP gates leading to the required ordering to only one side of the MPO.
This has the disadvantage that the MPO for (\ref{eqn:qftl}) must be calculated first, which is computationally intractable for large numbers of qubits.
A better approach is to apply a swap gate to the input qubits whenever one is applied to the output qubits.
This makes the ordering of the input qubits the same at all times as that of the output qubits.
This approach also produces the required ordering and invites an interpretation that the resulting MPO contains only interesting correlations rather than expensive swap correlations.

\begin{figure}[htb]
  \centering
  \includegraphics[scale=1.0]{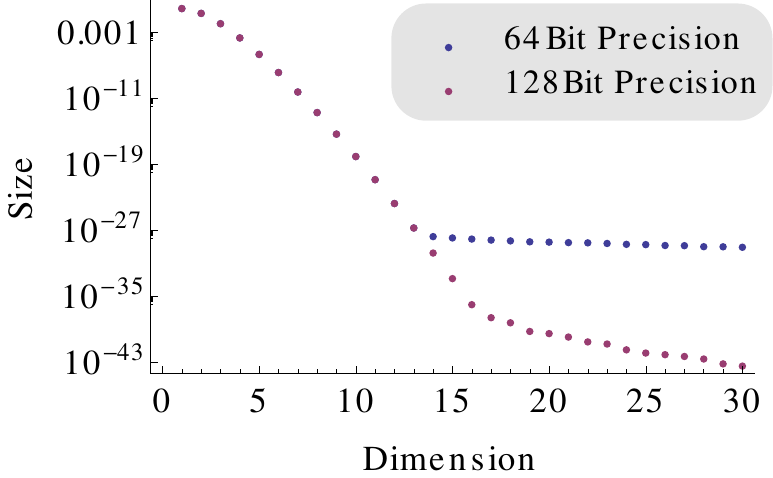}
  \caption{The probability distribution derived from the singular values of a bipartition at the center of MPOs representing the QFT with 25 qubits at two different precisions.}
  \label{fig:dimssize}
\end{figure}

We constructed MPOs encoding equation (\ref{eqn:qftlr}) with a simple nearest neighbour circuit \cite{Fowler2004}.
The bond ranks of the MPO encoding (\ref{eqn:qftlr}) were much lower than those required to encode (\ref{eqn:qftl}).
Figure \ref{fig:dimssize} shows the size of each element in a bond in the center of a MPO representing (\ref{eqn:qftlr}) for 24 qubits.
We display the probability distribution derived from the singular values $p_i = s_i^2 / D$ where $s_i$ are singular values and $D$ is the dimension of the Hilbert space ($2^{24}$ in this case).


The sizes of the bond elements displays a characteristic drop-off from lower to higher rank.
This characteristic was present regardless of the size of the MPO (MPOs with up to 50 qubits were tested).
Initially the rate of decrease is slow, but it quickly becomes exponential with the bond rank included.
The exponential decrease of bond element size halted at a probability of around $10^{-40}$ at quadruple precision (128 bits), which corresponds to a singular value of size relative to the largest value of $10^{-20}$.
There were many additional bond elements of this size or slightly smaller displaying a large amount of random variation in each MPO.
While their size is much larger than machine precision (these results were produced for quadruple precision numbers with $\epsilon \approx 10^{-35}$), the condition number of a singular value decomposition in this problem is very large and so instability at small element sizes is likely to result.
Additionally, computing the operator with a lower numerical precision (double precision with 64 bits for example) leads to a curve with the same exponential dropoff initially, but with the dropoff halting at a larger size.
It is thus likely that these elements are a result only of numerical imprecision.


The exponential dropoff of probability distribution values shown in figure \ref{fig:dimssize} has the implication that the Schmidt strength of the rearranged QFT converges to a constant value as the number of qubits in the transform is increased.
The Schmidt strength is the maximum entropy of the probability distribution following from the singular values of an operator $U$ along any bipartition.
It also gives the maximum entropy $E(U \ket{\alpha} \ket{\beta})$ where $\ket{\alpha}$ and $\ket{\beta}$ are states in two different quantum systems corresponding to any bipartition of the operator $U$.
The states $\ket{\alpha}$ and $\ket{\beta}$ are maximally entangled with ancillary systems \cite{Nielsen2003}.
We compute this strength to be $0.8208$.
For comparison, the Schmidt strength of a CNOT or CPHASE gate is $1$ and the Schmidt strength of a SWAP gate is $2$.

\begin{figure}[htb]
  \centering
  \includegraphics[scale=0.8]{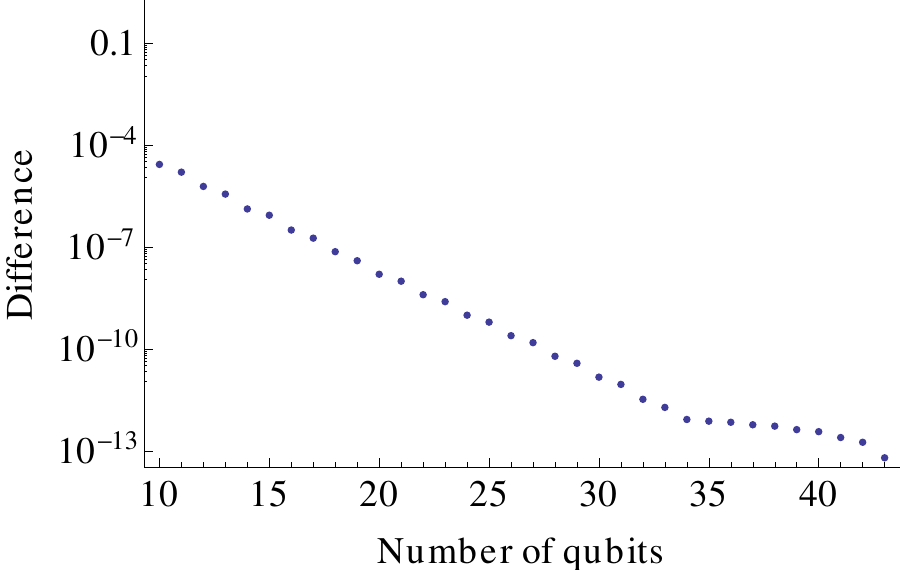}
  \caption{The mean difference between the probability distribution of the middle bond of the QFT computed with each number of qubits and that computed with the largest number of qubits (44).}
  \label{fig:dimconv}
\end{figure}

The convergence of the bond sizes is shown in figure \ref{fig:dimconv} where we plot the mean difference between the size values obtained with a given number of qubits and those of the largest MPO created (44 qubits).
It is clear that the values are converging towards the characteristic visible in figure \ref{fig:dimssize}a.
Note that for reasons of speed these results were computed at double precision and so the halting of the convergence at 34 qubits represents the calculation reaching machine precision.

\begin{figure}[htb]
  \centering
  \includegraphics[scale=0.8, clip, trim = 0 0 0 0]{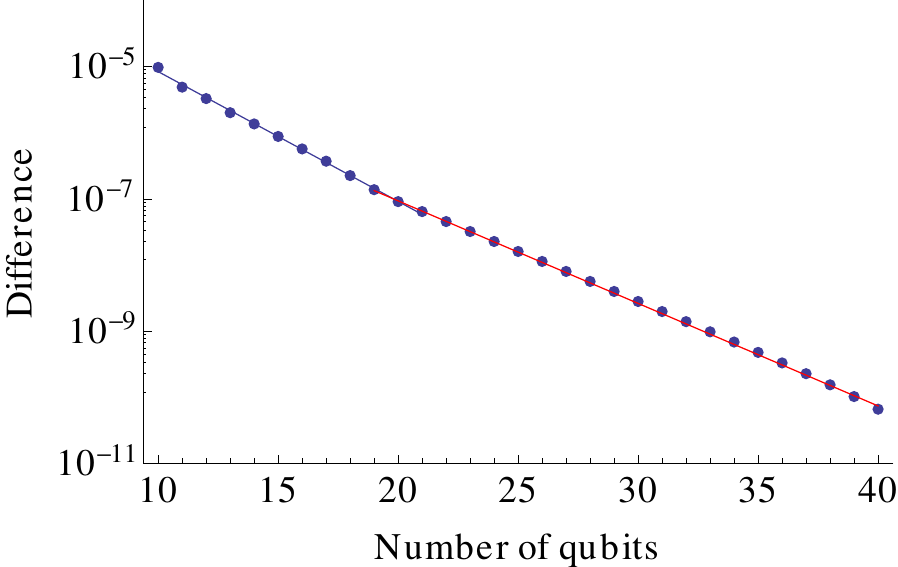}
  \caption{The mean difference between the size of the values in a tensor in the middle of a MPO and those of a tensor in the middle of a MPO of the largest size (44 qubits).
  These differences are normalised by the size of the maximum value in the tensor.
Two different decay rates are shown.}
\label{fig:matdiffs}
\end{figure}

After truncation of the smaller bond elements in the MPOs encoding (\ref{eqn:qftlr}), the tensors in the middle of the transform converged to a constant tensor as the number of qubits was increased.
This convergence completely specifies tensors in the middle of the transform up to phase rotations which result from the lack of uniqueness of the SVD, which can be easily corrected.
The convergence is illustrated in figure \ref{fig:matdiffs}, which shows the mean difference between the absolute values of the elements of the middle tensor of each MPO and the absolute values of the elements of middle tensor of the largest MPO computed (44 qubits).
Again, these results were computed at double precision.

Two different exponential decay rates are visible in the plot.
The first of these rates is the region at which we must truncate the middle tensor of the largest MPO to compare it to smaller tensors in smaller MPOs while from 20 qubits onwards, the truncation occurred at a bond size of 30 during calculation of the MPO and so the tensors are the same size.

It is clear from our results that the sizes of each bond in a MPO of the QFT decrease exponentially with increasing bond size.
Truncating a bond to size $t$ would thus create an error of $O(e^{-t})$, and $O(n)$ truncations in a $n$ qubit transform would create errors of size $O(n e^{-t})$.
To maintain a constant error as the number of qubits is increased, the bond size required would thus be $O(\log(n))$.
The convergence towards a common tensor in the middle of the transform implies that it would be possible to find a standard MPO form for the QFT with a relatively small number of qubits determined by a given error tolerance.
The middle tensor of this standard QFT could then be replicated a number of times to apply the transformation to any required larger number of qubits.
With a $n$ qubit QFT applied to a MPO with maximum Schmidt rank $\chi$, this would allow simulation of the QFT in $O(n \left(\log(n) \chi\right)^2)$ time.


\section{Truncation Errors}
\label{sec:trunerror}

Truncation of bonds of even small size will necessarily introduce error into the representation of an operator.
In order to confirm that an efficient MPO simulation of the QFT can be run with the bond size scalings suggested by figure \ref{fig:dimssize} without compromising accuracy, it is necessary to quantify this error.
It is difficult to quantify the error in a large MPO because the computational cost of calculating any interesting metric will in general grow exponentially with the number of qubits.
This is true of any calculation which does not take advantage of the structure of the MPO.
For example, many matrix norms require the calculation of the eigenvalues or decompositions of the full matrix of an operator, or maximisation of a function defined on the full matrix.
The matrix representation of the QFT is not sparse, and so the exact calculation of such quantities is intractable.

\begin{figure}[hbt]
  \centering
  \includegraphics[scale=0.85, clip, trim = 0 0 0 0]{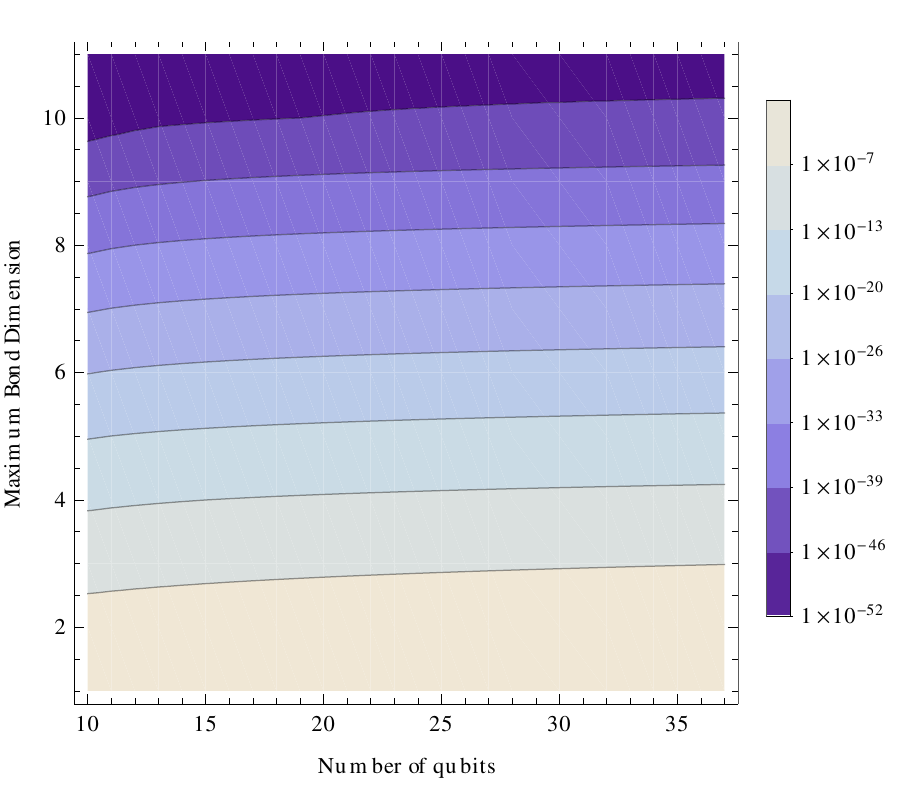}
  \caption{The error in the trace inner product between two MPOs of the QFT, one with truncation and one without any truncation.}
\label{fig:trinprod}
\end{figure}

Instead, we have calculated two less rigorous but more easily computed norms.
In order to perform the calculations at large system sizes, these calculations had to be performed with double precision.

Firstly, we computed the Hilbert-Schmidt inner product $\frac{1}{D}\text{tr}(U V^*)$ where $U$ is a MPO representing the QFT and whose bonds are truncated to a given size, $V$ is the same operator but is not truncated and $D$ is the dimension of the Hilbert space.
The value obtained measures the inner product between $U \ket{\psi}$ and $V \ket{\psi}$ averaged over all states $\ket{\psi}$.
The error in the result $1 - \frac{1}{D}\text{tr}(U V^*)$ is shown in figure \ref{fig:trinprod}.
It can be seen from this calculation that the error drops off exponentially as the bond rank is increased, and increases sub-exponentially as the number of qubits is increased.
However, this regime only extends as far as a maximum bond rank of 8, after which the observed error was zero.
At these ranks, the average error is thus below the machine precision of around $10^{-16}$.
It is worth noting that this is only an average measure of error and so does not reflect the worse case error involved in applying a truncated MPO.

The second measurement of error we computed is the amount of error associated with Fourier transforming a periodic state. A periodic state with $L$ qubits and period $r$ takes the form $\sum_{n=0}^{2^L/r-1} \ket{k_0 + nr}$.
These states are produced by the modular exponentiation stage of Shor's algorithm.
Applying the QFT to a periodic state produces a state which is strongly peaked around the values $\ket{i/r \, 2^L}$ for $i<r$ and so measuring the Fourier Transform of a periodic state reveals the period.
This is the basis of Shor's algorithm.

\begin{figure*}[hbt]
  \centering
  \subfloat[]{\includegraphics[scale=0.85]{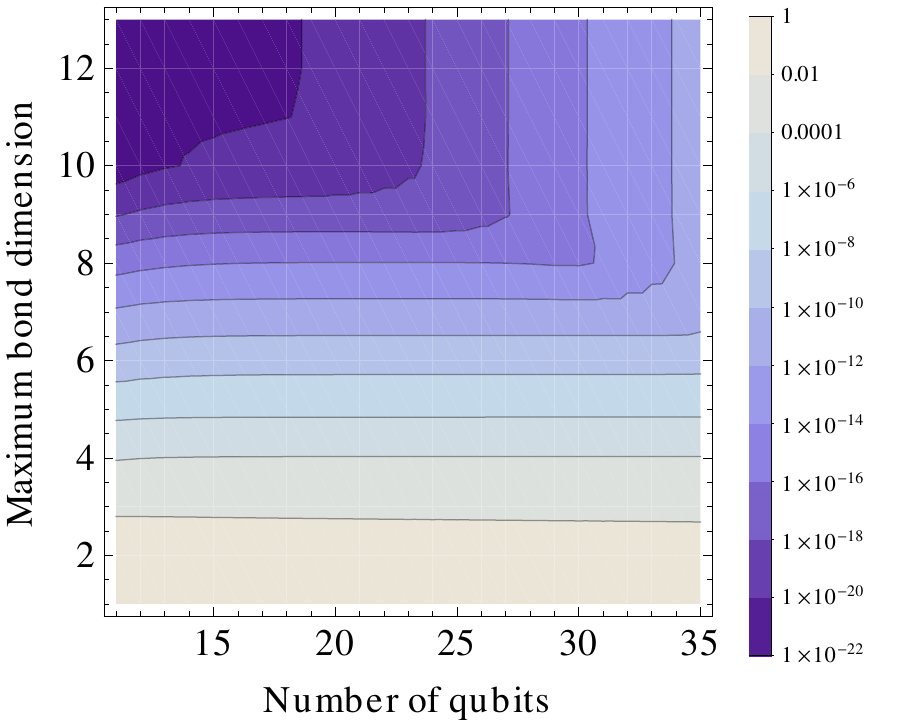}}\\
  \caption{The difference between the probability of measuring a peak after simulating a QFT with a MPO on a periodic state and the analytic probability. Shown for a period of 9.}
\label{fig:periodic}
\end{figure*}

We prepared periodic states with a range of periods and numbers of qubits between 10 and 28 and computed the deviation of the sizes of the peaks from the analytic values.
These results are shown in figure \ref{fig:periodic} for 9 qubits.
The results were similar for other periods for other values tested ($2\leq r\leq 15$).
As with the trace inner product, the error seems to decrease exponentially as the bond rank is increased at low bond ranks.
At higher bond ranks, the error appears to increase quickly as the number of qubits is increased.
It is difficult to obtain data with larger numbers of qubits due to the exponential scaling of the calculation of the analytic size of the peaks.

The results in figure \ref{fig:dimssize} indicate that many extra bond elements appear at double precision with sizes relative to the largest element of $10^{-13}$ or less.
We would expect that errors observed after simulating the QFT of periodic inputs would be at less than or equal to these levels.
This is the case for the range of qubits tested.

\section{Reasons for the efficient representation}
The fact that ordering the input values of the qubits differently to the output values can lead to a dramatic reduction in MPO complexity raises the question of whether a different ordering to that considered in (\ref{eqn:qftlr}) may be optimal.
We tested this by constructing MPOs with all possible input qubit orderings for QFTs with up to twelve qubits.
In every case the ordering in (\ref{eqn:qftlr}) was optimal.
For the reasons described above (the output at the nth qubit depends only upon the input at the first n qubits), we expect this to be the case for larger numbers of qubits as well.

\begin{figure}[hbt]
 \centering
 \label{fig:qftlb}
 \mbox{
\Qcircuit @C=0.3em @R=.5em {
  & \gate{H} & \ctrl{1} & \ctrl{2} & \qw & \qw & \qw & \qw & \qw & \qw & \qw & \qw & \qw & \qw\\
  & \qw & \gate{R_2} & \qw & \gate{H} & \ctrl{1} & \ctrl{2} & \qw & \qw & \qw & \qw & \qw & \qw & \qw\\
  & \qw & \qw & \gate{R_3} & \qw & \gate{R_2} & \qw & \gate{H} & \ctrl{1} & \ctrl{2} & \qw & \qw & \qw & \qw\\
  & \qw & \qw & \qw & \qw & \qw & \gate{R_3} & \qw & \gate{R_2} & \qw & \gate{H} & \ctrl{1} & \qw & \qw\\
  & \qw & \qw & \qw & \qw & \qw & \qw & \qw & \qw & \gate{R_3} & \qw & \gate{R_2} & \gate{H} & \qw
}
}
\caption{The AQFT for five qubits with a maximum of three controlled phase gates.}
\end{figure}
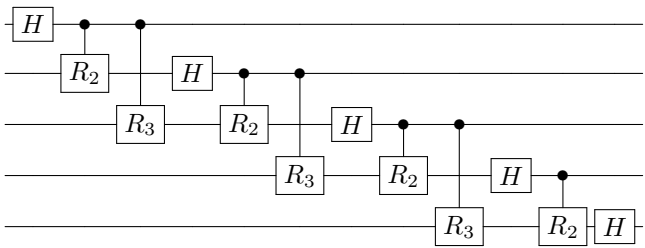

It would seem natural to explain the exponential decrease of bond element size shown in figure \ref{fig:dimssize} with the small effect of the rotation gates correlating far away qubits.
That is, the full QFT introduces correlations across every qubit pairing.
However, these correlations take the form of controlled phase rotations and the size of the rotations decreases inverse-exponentially with the one-dimensional distance between qubits.
As such, we should be able to neglect long range correlations.
We would expect this to cause the sizes of the tensors at the qubits within the MPO to be almost entirely unaffected by the number of far-away qubits.
This is the idea behind the AQFT \cite{Coppersmith2002}, where the number of controlled phase gates conditioned upon each qubit, henceforth the bandwidth, is set at a fixed value irrespective of the number of qubits in the transform.

\begin{figure}[htb]
  \centering
  \includegraphics[scale=0.65, clip, trim = 0 0 0 0]{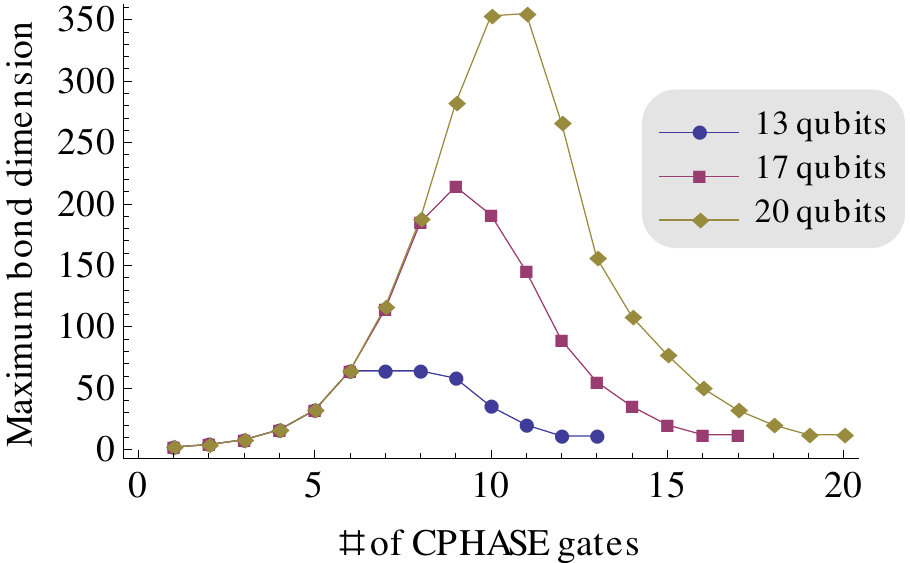}
  \caption{The maximum bond rank in MPOs corresponding to AQFTs with different numbers of qubits. Each AQFT was constructed with a different maximum number of controlled phase gates conditioned on each qubit.}
\label{fig:maxdims}
\end{figure}

However, we constructed MPOs using a nearest neighbour quantum circuit of the AQFT and found that the bond ranks produced were larger than those produced for the full transform.
The maximum bond ranks for a series of AQFTs after truncation are shown in figure \ref{fig:maxdims}.
Maximum bond ranks in an AQFT increased by a factor of 2 per additional controlled phase rotation included.
This increase levelled off in the middle of the transform but still quickly became computationally intractable.
Furthermore, the trace inner product between an operator truncated at any bond rank and a series of operators with reduced bandwidths was a maximum for the full transform and decreased monotonically as the bandwidth decreased.
It thus seems that the low required bond ranks observed in (\ref{eqn:qftlr}) are a feature of the full QFT.

While the low required bond rank of the QFT cannot be attributed entirely to decreasing phase rotations, the size of these rotations and the rate of their decrease are important.
We found that transforms with the same structure as the QFT but with phase rotations decreasing as $\exp \left(2 \pi i / k^n \right)$ instead of $\exp \left(2 \pi i / 2^k \right)$, where $k$ is the qubit distance and $n$ an integer, did not display the characteristic dropoff of bond size.
Rather, the required bond ranks appeared to increase with increasing numbers of qubits, presumably until the phase rotations become smaller than machine precision.
Using a rotation with some randomness in the form of $\exp \left(2 \pi i / 2^{k+\delta} \right)$ or $\exp \left(2 \pi i / (2+\delta)^{k} \right)$, with $\delta$ a small random number, also removed the exponential dropoff.
Rotations of the form $\exp \left(2 \pi i / n^k \right)$ for $n \geq 2$ still lead to Fourier transforms, although not over $\mathbb{Z}_{2^m}$, and were found to still lead to an exponential dropoff in bond element size. 
The rate of this dropoff increased as $n$ increased.

As such, while the small bond rank required to accurately represent the QFT with a MPO is not due solely to the decreasing size of the phase rotations used, it is related to them.
It is likely the exponential dropoff of bond size is the result of a symmetry in the structure of the QFT.
In order to obtain a low bond rank it is necessary to have phase rotations which decrease at least exponentially with the distance between qubits and to have the same phase rotation for each conditioned gate at a given linear qubit distance.

\section{Discussion}
\label{sec:disc}
While we have not proven that the QFT can be efficiently represented as a MPO, our numerical results are strongly suggestive of this.
If appears that the numerical error associated with the very small amount of truncation required for a tractable representation is very close to zero over a range of numbers of qubits.
Additionally, the differences between a matrix in the middle of each operator and an adjacent matrix decreases as the system size increases.

Together, these results suggest that a MPO representing a QFT for an arbitrary number of qubits can be created from the MPO representation of a QFT of a smaller size.
With an appropriate bond rank, this would allow the QFT to be performed on a MPO with maximum Schmidt rank $\chi$ with computational cost $O(n \left(\log(n)\right)^2 \chi^2)$.
It could similarly be performed on weakly correlated mixed states.
Our method allows the QFT to be efficiently simulated in a straightforward fashion in any case in which the qubits are ordered linearly.

Application of the QFT to a MPS of $n$ qubits with this method increases the bond rank by at most a factor scaling as $O(\log(n))$.
Denoting this factor by $d$, the application of $m$ QFTs increases the bond ranks by a maximum factor of $d^m$.
As such, the application of a constant number of QFTs can be efficiently simulated with a large number of qubits.
These QFTs can be interspersed by quantum circuits that do not increase the Schmidt rank.

Our results strengthen earlier work.
In \cite{Yoran2007a} the AQFT is show to be classically simulatable in polynomial time, although an explicit scaling is not derived.
Our method of simulation uses the full QFT and has a more advantageous scaling with respect to the number of qubits of $O(n \left(\log(n)\right)^2)$.

In \cite{Yoran2007a} a condition is also derived for when two efficiently simulatable quantum circuits composed may be efficiently simulated.
From this condition it follows that any circuit composed of a constant number of AQFTs and log-depth limited interaction range circuits can be efficiently classically simulated.
We provide a different perspective on the composability criteria.
That is, our method makes explicit the scaling of the cost of the QFT with the Schmidt rank of the bipartitions in the input state.
We have shown the difficulty of simulating the QFT to be mostly determined by the complexity of the state being transformed.
A log-depth limited interaction circuit will produce an input state with small Schmidt ranks across each bond partition, and so the previous result follows from our results.

That the QFT can be represented to very high fidelity with a MPO with limited bond ranks implies that the QFT can produce only a limited amount of entanglement.
This conclusion was originally shown in \cite{Yoran2008}, however our methods are more straightforward.


With respect to the question of where the quantum speedup in Shor's algorithm originates, our results provide further evidence that it originates in the highly entangled state generated by modular exponentiation.
Periodic states are generated by modular exponentiation, and a state of period $r$ will have bond ranks in a MPS of $r$.
As the maximum period of a modular exponentiation process factoring a number $N$ scales as $O(N)$ \cite{Nam2013}, states with very high Schmidt numbers are generated.
These states are very difficult to represent in a MPS and thus are very difficult to Fourier transform.
This conclusion is similar to that reached in other works such as \cite{Yoran2007a, Abbott2012}.
It is additionally worth noting that while our method makes very clear the connection between the Schmidt rank of the input state and the difficulty in Fourier transforming it, the same conclusion can be drawn about the computational speedup from the results of \cite{Yoran2007a}.

\section{Acknowledgements}
This research was conducted by the Australian Research Council Centre of Excellence for Quantum Computation and Communication Technology (project number CE110001027).

\bibliography{refmpopaper}

\end{document}